\documentclass[conference]{IEEEtran}
\IEEEoverridecommandlockouts
\usepackage{cite}
\usepackage{amsmath,amssymb,amsfonts}
\usepackage{graphicx}
\usepackage[dvipsnames, table,xcdraw]{xcolor}
\usepackage{xurl}
\usepackage{multirow}
\usepackage{makecell} 
\usepackage{booktabs}

\def\BibTeX{{\rm B\kern-.05em{\sc i\kern-.025em b}\kern-.08em
    T\kern-.1667em\lower.7ex\hbox{E}\kern-.125emX}}
\begin{document}

\title{Cross-Country Learning for National Infectious Disease Forecasting Using European Data}

\author{
	\IEEEauthorblockN{
	    Zacharias Komodromos\textsuperscript{a}, Kleanthis Malialis\textsuperscript{a}, Artemis Kontou\textsuperscript{a}, and Panayiotis Kolios\textsuperscript{a, b}
	}
	\IEEEauthorblockA{
		\textsuperscript{a} \textit{KIOS Research and Innovation Center of Excellence}\\
		\textsuperscript{b} \textit{Department of Computer Science}\\
		\textit{University of Cyprus},
		Nicosia, Cyprus
        \thanks{Email:\texttt{\{komodromos.zacharias, malialis.kleanthis, kontou.artemis, kolios.panayiotis\}@ucy.ac.cy}}
        \thanks{ORCID: 0009-0002-8999-1201 (ZK), 0000-0003-3432-7434 (KM), 0000-0003-0480-0796 (AK), 0000-0003-3981-993X (PK)}
        \thanks{This work was supported by the European Union’s Horizon 2020 research and innovation programme under grant agreement No 739551 (KIOS CoE - TEAMING) and from the Republic of Cyprus through the Deputy Ministry of Research, Innovation and Digital Policy. It was also supported by the CIPHIS (Cyprus Innovative Public Health ICT System) project of the \mbox{NextGenerationEU} programme under the Republic of Cyprus Recovery and Resilience Plan under grant agreement C1.1l2.}
         \thanks{\textcopyright\ 2026 IEEE. Personal use of this material is permitted. Permission from IEEE must be obtained for all other uses, in any current or future media, including reprinting/republishing this material for advertising or promotional purposes, creating new collective works, for resale or redistribution to servers or lists, or reuse of any copyrighted component of this work in other works.}
        }
}

\maketitle

\begin{abstract}
Accurate forecasting of infectious disease incidence is critical for public health planning and timely intervention. While most data-driven forecasting approaches rely primarily on historical data from a single country, such data are often limited in length and variability, restricting the performance of machine learning (ML) models. In this work, we investigate a cross-country learning approach for infectious disease forecasting, in which a single model is trained on time series data from multiple countries and evaluated on a country of interest. This setting enables the model to exploit shared epidemic dynamics across countries and to benefit from an enlarged training set. We examine this approach through a case study on COVID-19 case forecasting in Cyprus, using surveillance data of European countries. We evaluate multiple models and analyse the impact of the lookback window length and cross-country `data augmentation' on multi-step forecasting performance. The results show that combining data from other countries can lead to consistent improvements over models trained solely on national data. Although the focus is on Cyprus and COVID-19, the framework and findings provide promising insights for infectious disease forecasting in settings with limited national data.

\indent \textit{Clinical Relevance}—This work uses only aggregated, publicly available COVID-19 indicators (daily reported infections), avoiding individual-level data. The proposed forecasting pipeline can assist healthcare planning and inform government decisions on non-pharmaceutical interventions (e.g., lockdowns, social distancing) during epidemic surges.
\end{abstract}


\section{Introduction}
\subsection{Motivation}
The recurring threat of infectious diseases can be viewed as an ongoing evolutionary battle between humans and pathogenic microorganisms \cite{zappa2009emerging}. These diseases can spread within populations through direct contact, environmental exposure, or vectors such as mosquitoes.

Through the centuries, humanity has repeatedly faced the emergence of infectious diseases, resulting in demographic, economic and societal consequences, including the Black Death (1346--1353) \cite{benedictow2004black}, the 1918 influenza pandemic \cite{taubenberger20061918}, the HIV/AIDS pandemic \cite{jones2008global}, and the 2009 H1N1 influenza \cite{scalera2009first}. In recent years, and more specifically over the last century, globalisation, including increased human mobility and highly connected transport networks, has made geographic containment more difficult, allowing outbreaks to spread internationally within short timeframes \cite{jones2008global, tatem2006global}.

On the other hand, developments in biomedical technology and healthcare have substantially improved our ability to mitigate epidemics. Vaccination remains a key tool and has been recognised as a valuable public health achievement due to its impact on preventing infectious diseases \cite{koppaka2011ten, who_vaccination_history}.

A prominent example is COVID-19 \cite{chan2020familial, shi2020overview}, which emerged late 2019 and rapidly became a major global crisis. It caused widespread hospitalisations that strained healthcare systems in many countries, resulted in substantial mortality, and large numbers of reported infections that continue to this day. Governments worldwide faced the crisis differently, with some adopting strict measures, like travel restrictions, and others being less stringent. The World Health Organization (WHO) Emergency Committee ended the Public Health Emergency of International Concern (PHEIC) on 5 May 2023 \cite{who2023}. COVID-19 taught us important lessons on how to face a global health emergency, yet gaps remain in readiness for future outbreaks.

\subsection{Related work} 
Infectious disease forecasting is approached using \textit{data-driven} methods, \textit{compartmental} epidemiological models, \textit{agent-based} models, and \textit{graph-based} learning \cite{dimitriou2025reconstructing}, with the first two being the most widely used. All of these have been applied to COVID-19 forecasting. To mitigate data scarcity, \textit{cross-country} learning has been adopted in various studies.

\par \textbf{Data-driven methods.} This category includes traditional statistical techniques and more advanced ML and DL models.

Simple traditional \textit{statistical models} such as ARIMA \cite{sahai2020arima} have been used to forecast infectious diseases. These approaches often perform well for short-term forecasting.

Classical \textit{ML models}, such as simple linear regression \cite{abolmaali2021comparative}, random forests \cite{galasso2022random}, and gradient-boosted trees (e.g., LightGBM, XGBoost \cite{fang2022application}), have been explored for case forecasting. \textit{Deep learning} approaches, including Long Short-Term Memory (LSTM), Gated Recurrent Units (GRU), and Convolutional Neural Networks (CNNs), have also been analysed \cite{dairi2021comparative}.
However, for country-level daily series, the limited amount of historical data can constrain deep models and increase overfitting risk, motivating careful evaluation and data-augmentation strategies.
Moreover, incorporating exogenous signals like non-pharmaceutical interventions and weather variables can further improve modelling \cite{komodromos2025investigating, yeung2021machine}.
Online incremental learning \cite{malialis2020online} has also been suggested to adapt models as data distributions evolve over time \cite{stylianides2023study}.

\par \textbf{Compartmental models.} These models partition the population into distinct `compartments' according to disease status. The compartment sizes evolve over time and are typically modelled using ordinary differential equations (ODE), which describe transitions over time. A widely used example is the Susceptible-Infected-Recovered (SIR) model. More complex extensions include the SIDAREVH model \cite{karapitta2024time, karapitta2024pandemic}, which accounts for vaccination status. These models are interpretable and can capture broad epidemic dynamics under explicit assumptions about transmission and progression.

\textbf{Cross-country/region learning.}
To combat data shortage, several studies incorporate information across different countries or regions.
A multi-task Gaussian Process that leverages both task-specific (individual) and cross-task (shared) information to predict cases and deaths has been studied \cite{kim2025covid}. Additionally, a model-agnostic meta-learning (MAML) region-adaptive framework for influenza forecasting was introduced to mitigate data limitations by embedding common prior knowledge from multiple tasks into the target model parameters \cite{moon2023model}. Finally, Epidemiology-Aware Neural ODE with Continuous Disease Transmission Graph (EARTH), a global-to-local approach that models transmission dynamics using graphs and global trends, has also been proposed \cite{wanearth}.

Our approach uses a single forecasting model trained on pooled cross-country time series for the country-specific target, quantifying the impact of incorporating additional data, without requiring multi-output, multi-task formulations or explicit transmission-graph modeling.

\subsection{Contributions}
The above motivate leveraging the large amount of information generated during the pandemic to improve decision-making and public health preparedness, including surveillance, outbreak control, and resource planning during surges of infectious diseases, potentially reducing health, social, and economic consequences. More precisely, accurate short-term forecasting can enhance government response by enabling earlier, better-targeted interventions.

The contributions of this work are:

\begin{itemize}
    \item A cross-country learning framework for infectious disease case forecasting is proposed, in which a single forecasting model is trained on time series data from multiple countries and evaluated on a country of interest. This setting enables it to exploit shared epidemic dynamics across countries and improves performance compared to models trained solely on national historical data.
    \item An extensive case study on COVID-19 infection forecasting for Cyprus is presented, using surveillance data from European countries. We evaluate ML forecasting methods and systematically analyse the impact of (i) the lookback (historical) window length and (ii) augmenting the training data with cross-country observations.   
    \item While the empirical focus is on Cyprus, the findings provide insights that are relevant to public health decision-makers and health experts in other European countries, particularly in cases of limited national historical data.
\end{itemize}

The paper is structured as follows. Section~\ref{sec:data_methods} describes the data, the cross-country learning setup, the compared models, and the evaluation method. Section~\ref{sec:results} reports and analyses the results, and finally, Section~\ref{sec:conclusion} concludes the paper.

\section{Data \& Methods}\label{sec:data_methods}
\subsection{Data}
\textbf{Dataset.} Our dataset consists of daily, country-level observations of SARS-CoV-2 confirmed cases. The data were obtained from the Oxford COVID-19 Government Response Tracker (OxCGRT) \cite{policies} and cover the period 2020-01-01 to 2022-12-31. For Cyprus, the confirmed cases were updated using public data from the Ministry of Health of the Republic of Cyprus, published throughout the pandemic in weekly reports by the Press and Information Office, as this reporting was more consistent than the corresponding OxCGRT series.

\begin{figure}[!htbp]
    \vspace{-5pt}
    \centering
    \includegraphics[width=0.49\textwidth]{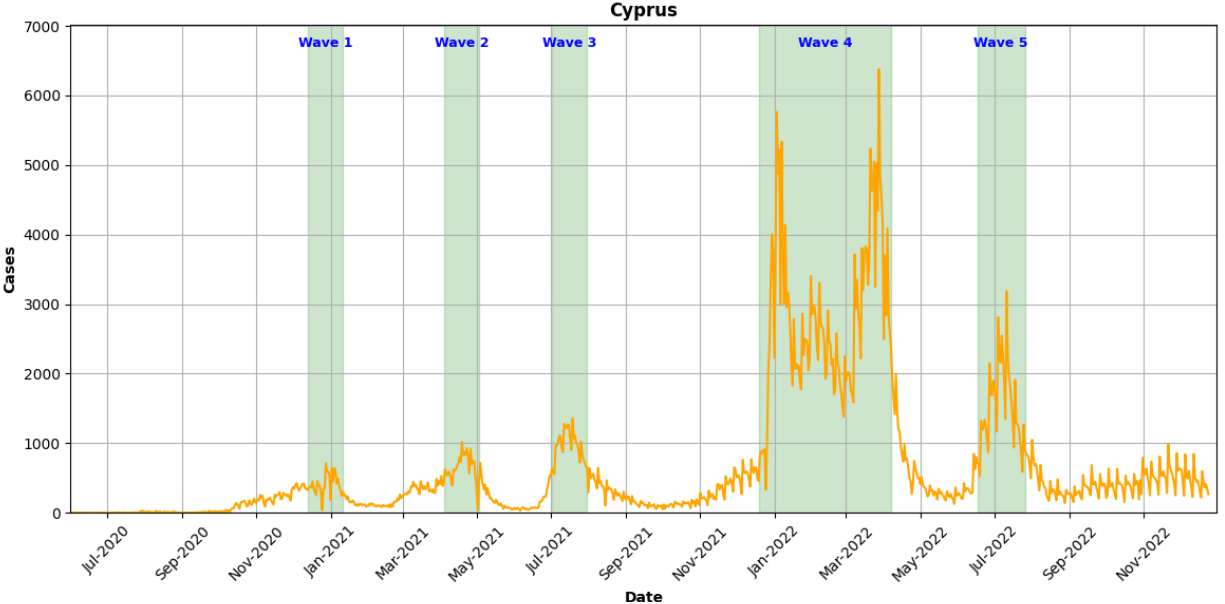}
    \caption{Cyprus' COVID-19 cases per day (Period: 2020-06-01 to 2022-12-31 for visual purposes).}
    \label{fig:cases_cyprus}
\end{figure}

During the study period, Cyprus experienced five distinct pandemic waves (Figure~\ref{fig:cases_cyprus}):
\begin{enumerate}
  \item 13 December 2020 to 11 January 2021,
  \item 4 April 2021 to 3 May 2021,
  \item 2 July 2021 to 31 July 2021,
  \item 19 December 2021 to 7 April 2022,
  \item 17 June 2022 to 26 July 2022.
\end{enumerate}

\textbf{Reporting across countries.} Reporting frequency differs across countries and over time, meaning that case counts were not always reported daily. Therefore, apparent `missing values' are often semi-missing, as we harmonise non-reporting days by distributing reported totals across the corresponding period to maintain consistency in the daily time series representation. For example, if a country reported weekly for a period, the weekly total was averaged and distributed evenly across the seven days. The dataset consists of 46 countries. As a rule of thumb, we restrict our country selection to those with fewer than 1/9 of reporting days missing.

\begin{figure*}[!htbp]
    \centering
    \includegraphics[width=\textwidth]{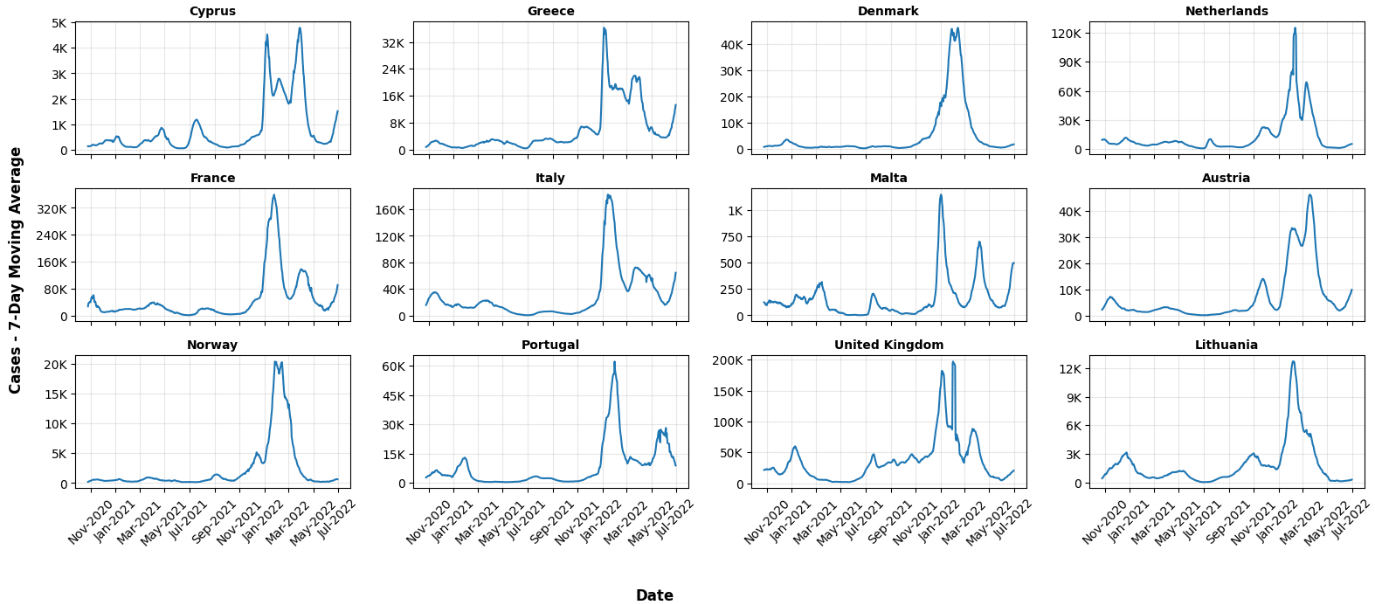}
    \caption{7-day moving averages of confirmed cases across countries.}
    \label{fig:top12countries_timeseries}
    \vspace{-8pt}
\end{figure*}

Figure~\ref{fig:top12countries_timeseries}, which includes Cyprus and the top 11 countries by Spearman correlation with Cyprus, shows 7-day moving averages of confirmed cases, highlighting a very high wave in early 2022. Among the selected countries, Greece exhibits the closest temporal alignment with Cyprus.

\textbf{Data pre-processing.} As mentioned, missing reporting days were interpolated by distributing the total count reported on a given date, across the corresponding non-reported days.

Then, log-transformation (log1p) was applied to handle extreme values and exponential growth, while reducing variance by compressing the range and addressing right-skewness.

For the Transformer model, we apply per-country standardisation to scale inputs to zero mean and unit variance, improving numerical stability and facilitating training. Statistical models are unaffected by scaling, and tree-based models such as XGBoost are largely insensitive because monotonic transformations preserve value ordering. However, when training on pooled multi-country data without an explicit country identifier, country-wise normalisation is used to account for cross-country differences in magnitude and to prevent high-incidence countries from dominating the learning signal.

\subsection{Cross-country learning}
We consider the problem of national infectious disease case forecasting from time series surveillance data.

\noindent \textbf{Single-country training.} For a country of interest $c^\ast$, we are given a historical time series:
\[
X^{(c^\ast)} = \{x_1^{(c^\ast)}, x_2^{(c^\ast)}, \dots, x_T^{(c^\ast)}\} \in \mathbb{N}_0^{\,T},
\]
where $x_t^{(c^\ast)} \in \mathbb{N}_0$ denote the reported number of positive cases at time $t$, and $T$ is the total number of days.

Using a lookback window of length $L$ and a forecasting horizon $h$, we construct supervised training pairs as follows:
\[
\mathbf{x}_t^{(c^\ast)} = \big[x_{t-L+1}^{(c^\ast)}, \dots, x_t^{(c^\ast)}\big], \quad
\mathbf{y}_t^{(c^\ast)} = \big[x_{t+1}^{(c^\ast)}, \dots, x_{t+h}^{(c^\ast)}\big].
\]
A native forecasting model is trained to learn the mapping:
\begin{equation}
f^{(c^\ast)} : \mathbf{x}_t^{(c^\ast)} \mapsto \mathbf{y}_t^{(c^\ast)}.
\end{equation}

\noindent \textbf{Cross-country training.}
Let $\mathcal{C} = \{c_1, \dots, c_C\}$ denote a set of countries, with $c^\ast \in \mathcal{C}$. For each country $c \in \mathcal{C}$, analogous input-output pairs $(\mathbf{x}_t^{(c)}, \mathbf{y}_t^{(c)})$ are constructed using the same lookback window $L$ and forecasting horizon $h$. The cross-country training dataset is defined as:
\[
\mathcal{D}_{\text{cross}} = \bigcup_{c \in \mathcal{C}} 
\left\{ \left(\mathbf{x}_t^{(c)}, \mathbf{y}_t^{(c)}\right) \right\}.
\]
A cross-country forecasting model $f_{\text{cross}}$ is trained on $\mathcal{D}_{\text{cross}}$ to learn shared epidemic dynamics across countries:
\begin{equation}
f_{\text{cross}} : \mathbf{x}_t^{(c)} \mapsto \mathbf{y}_t^{(c)}, \quad \forall c \in \mathcal{C}.
\end{equation}
The selection of countries is examined in Subsection~\ref{subsec:eval_method}.

\noindent \textbf{Cross-country inference.}
At inference time, both models are evaluated on the same target country $c^\ast$. In particular, the cross-country model produces forecasts:
\begin{equation}
\hat{\mathbf{y}}_t^{(c^\ast)} = f_{\text{cross}}\!\left(\mathbf{x}_t^{(c^\ast)}\right),
\end{equation}
allowing a direct comparison between native and cross-country training under identical forecasting targets.

\subsection{Compared models}
We compare our forecasting models against baseline methods that are widely used in forecasting and often prove surprisingly competitive, providing strong benchmarks that can be difficult to beat in practice \cite{hyndman2018forecasting}.

\textbf{Transformers} \cite{vaswani2017attention} are a class of neural network architectures designed to model sequential and temporal dependencies by relying on self-attention mechanisms, rather than recurrence. By computing pairwise interactions between all elements of an input sequence, Transformers are able to capture long-range dependencies efficiently and in parallel, making them suitable for time series forecasting.

\textbf{XGBoost} (eXtreme Gradient Boosting) \cite{chen2016xgboost} is a fast ML method based on gradient-boosted decision trees that constructs an ensemble model by iteratively adding decision trees. At each iteration, a new tree is trained to minimise a differentiable loss function by fitting the residual errors produced by the current ensemble. 

\textbf{ARIMA} (Autoregressive Integrated Moving Average) \cite{box2013box} is a classical statistical baseline method for univariate time series forecasting. It is often strong in short-term prediction, and has been widely used for COVID-19 case forecasting due to its simplicity and interpretability \cite{sahai2020arima}. In our experiments, the order was selected separately for each split using training data only, and choosing the one with the lowest Akaike Information Criterion (AIC) among fits that converged.

\textbf{Other baseline methods.} Let $y_t$ denote the observed daily value at day $t$. For a forecast issued at origin time $t$, the $h$-step-ahead prediction is denoted $\hat{y}_{t+h}$.

\begin{enumerate}
  \item \textbf{Na\"ive last (last value).} The future is assumed equal to the most recent observation:
  \begin{equation}
    \hat{y}_{t+h} = y_t,\qquad h=1,2, \dots, 7.
  \end{equation}
  \item \textbf{Last week's average.} The forecast equals the mean of the seven most recent observations:
  \begin{equation}
    \hat{y}_{t+h} = \frac{1}{7}\sum_{i=0}^{6} y_{t-i},\qquad h=1, 2, \dots, 7.
  \end{equation}
  \item \textbf{Seasonal na\"ive (weekly).} For daily data with weekly structure ($s=7$), each future day is predicted by the observation from the same weekday one week earlier:
  \begin{equation}
    \hat{y}_{t+h} = y_{t+h-s},\qquad s=7,\;\; h=1,2, \dots, 7.
  \end{equation}
\end{enumerate}

\subsection{Evaluation method and performance metrics}\label{subsec:eval_method}

\textbf{Country-selection categories.}
The selection of countries used in the models is based on two criteria. The number of missing reporting days, and Spearman's rank correlation between European countries and Cyprus. We first remove countries with more than one ninth of reporting days missing during the training period. We then divide the remaining countries into four categories (using the training period):
\begin{enumerate}
  \item National (only Cyprus),
  \item Correlated with Cyprus (Spearman $|\rho| \ge 0.3$),
  \item Uncorrelated with Cyprus (Spearman $|\rho| < 0.3$),
  \item All countries.
\end{enumerate}

\begin{figure*}[!htbp]
    \centering
    \includegraphics[width=\textwidth]{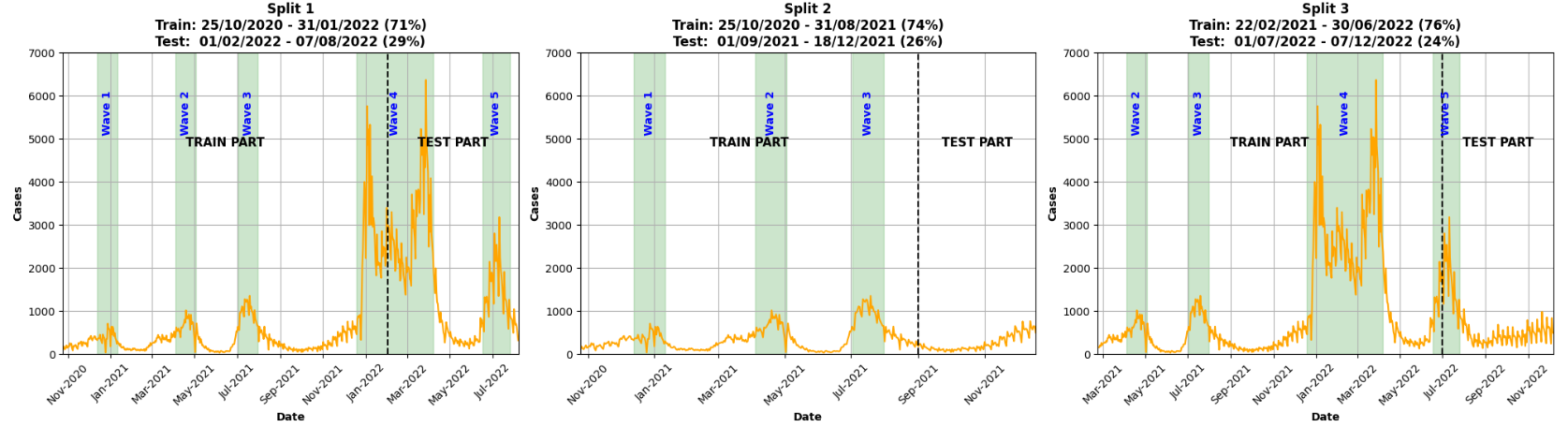}
    \caption{Train–test splits used in the experiments.}
    \label{fig:splits}
    \vspace{-5pt}
\end{figure*}

\textbf{Train-test splits.} For our study three train-test splits have been used (Figure~\ref{fig:splits}). In the first split, the training set includes the first half of the largest wave, while the test set covers the remainder of that wave and the final wave. This allows evaluation during both waves while retaining partial information from the largest wave. The second split includes three smaller waves in the training set and evaluates performance during low-activity periods, as the test set contains no wave days. The third split uses data from the later stages of the study period to assess performance in the final part of the series.

\textbf{Performance metrics}. \textbf{\textit{MAPE.}} The Mean Absolute Percentage Error (MAPE) is used as the primary evaluation metric, because it normalises errors, making performance comparable across periods with very different case levels.

\vspace{-5pt}
\begin{equation}
\text{MAPE}_d = \frac{100\%}{n} \sum_{k=1}^{n} \left| \frac{y_{k,d} - \hat{y}_{k,d}}{y_{k,d}} \right|,
\label{eq:mape}
\end{equation}
\vspace{-5pt}

\noindent where \(y_{k,d}\) is the true value for the \(k\)\textsuperscript{th} observation at the \(d\)\textsuperscript{th} day ahead forecast, \(\hat{y}_{k,d}\) is the corresponding predicted value, and \(n\) is the total number of predictions. No zero-valued observations occurred in the evaluated test periods (if present, they were replaced by 1).

\textbf{\textit{MAE.}} The Mean Absolute Error (MAE) is additionally reported as a direct, scale-dependent measure of forecast error. 
\begin{equation}
\text{MAE}_d = \frac{1}{n} \sum_{k=1}^{n} \left| y_{k,d} - \hat{y}_{k,d} \right|.
\label{eq:mae}
\end{equation}

Our models produce 7-day-ahead multi-step forecasts, so performance is summarised across the entire horizon. The averaged error metrics are defined as follows:
\begin{equation}
\text{MAPE}_{\text{7-day average}} = \frac{1}{7} \sum_{d=1}^{7} \text{MAPE}_d ,
\label{eq:mape_7days_avg}
\end{equation}
where \(\text{MAPE}_d\) is the MAPE of the \(d\)\textsuperscript{th} day ahead (see eq.~\eqref{eq:mape}).
The corresponding \(\text{MAE}_\text{7-day average}\) is defined similarly, by averaging the daily MAEs over the forecasting horizon.

We then compare the sum of predicted values against the sum of actual values to aggregate the results.

\begin{equation}
\begin{split}
\text{MAPE}_{\text{7-day aggreg.}} &= \frac{100\%}{n} \sum_{k=1}^{n}
\left| \frac{\sum_{m=1}^{7} y_{k,m} - \sum_{m=1}^{7} \hat{y}_{k,m}}
{\sum_{m=1}^{7} y_{k,m}} \right|,
\end{split}
\label{eq:mape_7days_aggr}
\end{equation}

\begin{equation}
\text{MAE}_{\text{7-day aggreg.}} = \frac{1}{n} \sum_{k=1}^{n} \left| \sum_{m=1}^{7} y_{k,m} - \sum_{m=1}^{7} \hat{y}_{k,m} \right|.
\label{eq:mae_7days_aggr}
\end{equation}

For the XGBoost and Transformer experiments, models were trained and evaluated over 15 repetitions. Reported results include the standard deviation of the evaluation metrics across repetitions in parentheses.

\section{Results \& Discussion}\label{sec:results}

\subsection{Role of the lookback window length (7/14/21 days)}

In this subsection, we examine how lookback window length affects forecasting by using 1, 2, and 3 weeks of history.

\begin{table*}[!htbp]
\centering
\caption{Results across different lookback windows (7/14/21 days).}
\vspace{-5pt}
\label{tab:lookback_window}
\small
\renewcommand{\arraystretch}{1.15}

\resizebox{\linewidth}{!}{%
\begin{tabular}{llcccccccccccc}
\toprule
\multirow{2}{*}{Model} & \multirow{2}{*}{\makecell{Countries\\Selection}}
& \multicolumn{4}{c}{7-Day Lookback}
& \multicolumn{4}{c}{14-Day Lookback}
& \multicolumn{4}{c}{21-Day Lookback} \\
\cmidrule(lr){3-6}\cmidrule(lr){7-10}\cmidrule(lr){11-14}
& &
\makecell{MAE\\Avg} & \makecell{MAPE\\Avg (\%)} & \makecell{MAE\\Aggreg.} & \makecell{MAPE\\Aggreg. (\%)} &
\makecell{MAE\\Avg} & \makecell{MAPE\\Avg (\%)} & \makecell{MAE\\Aggreg.} & \makecell{MAPE\\Aggreg. (\%)} &
\makecell{MAE\\Avg} & \makecell{MAPE\\Avg (\%)} & \makecell{MAE\\Aggreg.} & \makecell{MAPE\\Aggreg. (\%)} \\
\midrule
\multirow{2}{*}{XGBoost} &
National  & \textbf{431 (8)} & 27.5 (0.4) & \textbf{2355 (37)} & \textbf{20.8 (0.2)} & 461 (9) & \textbf{27.4 (0.5)} & 2671 (47) & 21.4 (0.3) &  465 (8) & 27.9 (0.6) & 2723 (38) & 21.8 (0.3) \\
& All Countries & 406 (4) & 28.3 (0.3) & 2408 (10) & 22.9 (0.1) & \textbf{361 (7)} & \textbf{24.5 (0.3)} & \textbf{2029 (21)} & \textbf{18.7 (0.1)} & 388 (7) & 26.0 (0.4) & 2135 (22) & 19.6 (0.2) \\

\midrule
\multirow{2}{*}{Transformer} &
National   & \textbf{443 (72)} & \textbf{30.7 (5.6)} & \textbf{2584 (544)} & \textbf{23.5 (5.2)} & 487 (109) & 32.4 (7.8) & 2854 (829) & 24.3 (7.4) & 492 (115) & 33.7 (8.8) & 2906 (923) & 25.7 (8.7) \\
& All Countries & 394 (60) & 26.1 (4.3) & 2362 (424) & 21.4 (3.9) & \textbf{370 (70)} & \textbf{24.1 (3.8)} & \textbf{2113 (509)} & \textbf{18.6 (3.6)} & 384 (102) & 24.8 (5.4) & 2215 (785) & 19.4 (5.6) \\
\bottomrule
\end{tabular}%
}
\vspace{-5pt}
\end{table*}

As shown in Table~\ref{tab:lookback_window}, the 14-day lookback window yields the best performance in the \textit{all countries} setting, while in the \textit{national} setting, the 7-day window has a slight edge over the other two windows. Given that the best results are achieved under the \textit{all countries} 14-day configuration across both models, we therefore adopt a two-week window, which provides a good balance between capturing short-term dynamics and avoiding unnecessary noise from longer lookback windows.

\subsection{Role of integrating other countries}

ML models, and particularly DL models, typically benefit from larger training sets. However, country-level daily COVID-19 series provide limited samples (365 observations per year per country). To increase the training data, we integrate the Cyprus training set with additional European countries and evaluate the impact on predictive performance.

\textbf{Correlation analysis.} For the figures and the analysis, we considered the three train-test splits (Figure~\ref{fig:splits}) and used the period spanning the earliest training start date to the latest training end date across splits (2020-10-25 to 2022-06-30).

\begin{figure}[!htbp]
    \vspace{-8pt}
    \centering
    \includegraphics[width=0.49\textwidth]{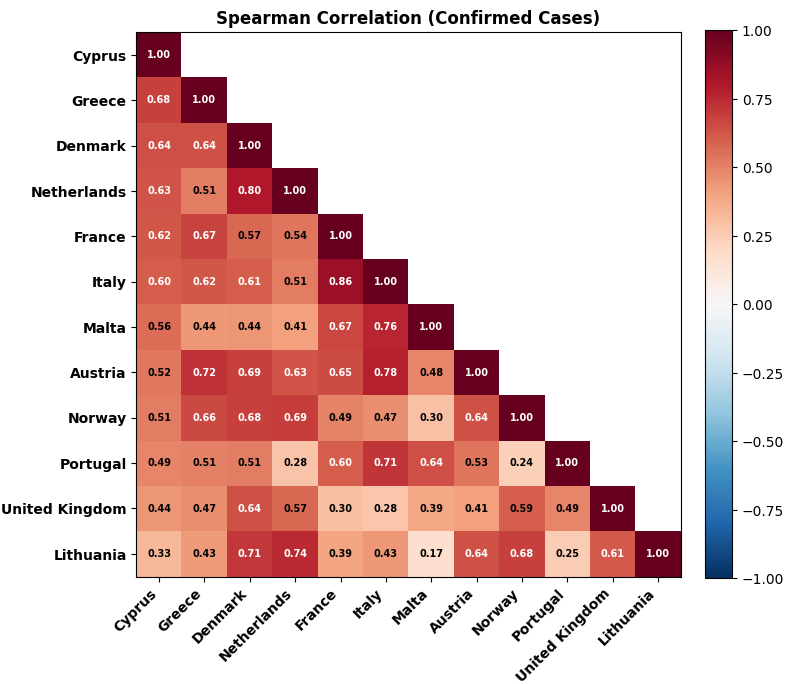}
    \caption{Spearman’s correlation matrix - Cases (Top 11 corr. with Cyprus).}
    \label{fig:spearmans_correlation}
    \vspace{-5pt}
\end{figure}

To quantify similarity between Cyprus and other European case trajectories over time, we compute Spearman's correlation \cite{spearman1961proof}, which captures monotonic association and is less sensitive to nonlinearity and extreme values than Pearson correlation. We report both Spearman and Pearson correlations (Table~\ref{tab:correlations}) and visualise the Spearman's matrix in Figure~\ref{fig:spearmans_correlation} (including Cyprus and the top 11 countries correlated with Cyprus). Greece exhibits the strongest association with Cyprus ($\rho=0.68$), while several countries show moderate correlations. Differences between Spearman and Pearson values suggest that some country pairs exhibit similar temporal ordering of waves even when the relationship is not strongly linear. Correlations along with missing counts guide our country-selection strategy in subsequent experiments.

\begin{table*}[t]
\centering
\caption{Correlation of daily cases with Cyprus (after preprocessing).}
\vspace{-6pt}
\label{tab:correlations}
\scriptsize
\setlength{\tabcolsep}{2.6pt}
\renewcommand{\arraystretch}{1.18}

\resizebox{\textwidth}{!}{%
\begin{tabular}{lccccccccccccccc}
\hline
\textbf{Country} &
\textbf{Greece} & \textbf{Denmark} & \textbf{Iceland} &
\makecell{\textbf{Nether-}\\\textbf{lands}} &
\textbf{France} & \textbf{Monaco} &
\makecell{\textbf{Luxem-}\\\textbf{bourg}} &
\textbf{Italy} &
\makecell{\textbf{Switzer-}\\\textbf{land}} &
\makecell{\textbf{Liechten-}\\\textbf{stein}} &
\textbf{Germany} & \textbf{Finland} & \textbf{Malta} & \textbf{Spain} & \textbf{Andorra} \\ \hline

\textbf{Spearman} &
\cellcolor[HTML]{63BE7B}0.68 & \cellcolor[HTML]{75C47D}0.64 & \cellcolor[HTML]{79C57D}0.63 &
\cellcolor[HTML]{7AC57D}0.63 & \cellcolor[HTML]{80C77D}0.62 & \cellcolor[HTML]{80C77D}0.62 &
\cellcolor[HTML]{84C87D}0.61 & \cellcolor[HTML]{87C97E}0.60 & \cellcolor[HTML]{88C97E}0.60 &
\cellcolor[HTML]{8ACA7E}0.60 & \cellcolor[HTML]{8BCA7E}0.59 & \cellcolor[HTML]{8BCA7E}0.59 &
\cellcolor[HTML]{99CE7F}0.56 & \cellcolor[HTML]{9ACE7F}0.56 & \cellcolor[HTML]{A3D17F}0.54 \\ \hline

\textbf{Pearson} &
\cellcolor[HTML]{63BE7B}0.90 & \cellcolor[HTML]{DDE283}0.58 & \cellcolor[HTML]{BDD881}0.67 &
\cellcolor[HTML]{E4E483}0.57 & \cellcolor[HTML]{C8DC81}0.64 & \cellcolor[HTML]{BAD780}0.68 &
\cellcolor[HTML]{B6D680}0.69 & \cellcolor[HTML]{B9D780}0.68 & \cellcolor[HTML]{9ACE7F}0.76 &
\cellcolor[HTML]{CCDD82}0.63 & \cellcolor[HTML]{ABD380}0.71 & \cellcolor[HTML]{9CCF7F}0.75 &
\cellcolor[HTML]{CFDE82}0.62 & \cellcolor[HTML]{E7E583}0.56 & \cellcolor[HTML]{FEEA83}0.49 \\ \hline\hline

\textbf{Country} &
\textbf{Austria} & \textbf{Norway} & \makecell{\textbf{San}\\\textbf{Marino}} &
\textbf{Belgium} & \textbf{Portugal} & \textbf{Ireland} &
\makecell{\textbf{United}\\\textbf{Kingdom}} &
\textbf{Lithuania} & \textbf{Turkey} & \textbf{Slovenia} & \textbf{Estonia} &
\makecell{\textbf{Czech}\\\textbf{Republic}} &
\textbf{Hungary} & \textbf{Latvia} & \textbf{Croatia} \\ \hline

\textbf{Spearman} &
\cellcolor[HTML]{A9D380}0.52 & \cellcolor[HTML]{B0D580}0.51 & \cellcolor[HTML]{B5D680}0.50 &
\cellcolor[HTML]{B7D780}0.49 & \cellcolor[HTML]{B9D780}0.49 & \cellcolor[HTML]{C4DA81}0.46 &
\cellcolor[HTML]{CDDD82}0.44 & \cellcolor[HTML]{FFEB84}0.33 & \cellcolor[HTML]{FEEA83}0.33 &
\cellcolor[HTML]{FEE783}0.32 & \cellcolor[HTML]{FEE582}0.31 & \cellcolor[HTML]{FEE081}0.29 &
\cellcolor[HTML]{FEDD81}0.28 & \cellcolor[HTML]{FDD880}0.26 & \cellcolor[HTML]{FDD27F}0.24 \\ \hline

\textbf{Pearson} &
\cellcolor[HTML]{A7D27F}0.72 & \cellcolor[HTML]{EAE583}0.55 & \cellcolor[HTML]{EEE784}0.54 &
\cellcolor[HTML]{FDEB84}0.50 & \cellcolor[HTML]{FFEB84}0.50 & \cellcolor[HTML]{C4DA81}0.65 &
\cellcolor[HTML]{E3E383}0.57 & \cellcolor[HTML]{F0E784}0.54 & \cellcolor[HTML]{FBA977}0.26 &
\cellcolor[HTML]{FEE182}0.46 & \cellcolor[HTML]{FEE883}0.49 & \cellcolor[HTML]{FCC27C}0.35 &
\cellcolor[HTML]{FBA977}0.26 & \cellcolor[HTML]{FCEA84}0.51 & \cellcolor[HTML]{FCBD7B}0.33 \\ \hline\hline

\textbf{Country} &
\textbf{Russia} & \textbf{Poland} & \textbf{Bulgaria} & \textbf{Sweden} & \textbf{Georgia} &
\textbf{Serbia} & \textbf{Romania} & \textbf{Kosovo} &
\makecell{\textbf{Bosnia}\\\textbf{\& Herz.}} &
\textbf{Ukraine} & \textbf{Moldova} &
\makecell{\textbf{Azer-}\\\textbf{baijan}} &
\textbf{Belarus} &
\makecell{\textbf{Kazakh-}\\\textbf{stan}} &
\textbf{Albania} \\ \hline

\textbf{Spearman} &
\cellcolor[HTML]{FDCF7E}0.23 & \cellcolor[HTML]{FDCE7E}0.22 & \cellcolor[HTML]{FDCC7E}0.22 &
\cellcolor[HTML]{FCC47C}0.19 & \cellcolor[HTML]{FCB379}0.13 & \cellcolor[HTML]{FBA376}0.07 &
\cellcolor[HTML]{FAA075}0.06 & \cellcolor[HTML]{FA9673}0.02 & \cellcolor[HTML]{FA9573}0.01 &
\cellcolor[HTML]{FA9272}0.00 & \cellcolor[HTML]{FA9172}0.00 & \cellcolor[HTML]{FA8E72}-0.01 &
\cellcolor[HTML]{F8776D}-0.10 & \cellcolor[HTML]{F8736D}-0.11 & \cellcolor[HTML]{F8696B}-0.15 \\ \hline

\textbf{Pearson} &
\cellcolor[HTML]{FDCC7E}0.39 & \cellcolor[HTML]{FBA777}0.25 & \cellcolor[HTML]{FCC07B}0.34 &
\cellcolor[HTML]{FDCB7D}0.38 & \cellcolor[HTML]{FCB679}0.31 & \cellcolor[HTML]{FBAE78}0.28 &
\cellcolor[HTML]{FBA977}0.26 & \cellcolor[HTML]{F97C6E}0.09 & \cellcolor[HTML]{F98370}0.12 &
\cellcolor[HTML]{F9826F}0.11 & \cellcolor[HTML]{F98C71}0.15 & \cellcolor[HTML]{F8696B}0.02 &
\cellcolor[HTML]{FA9A74}0.20 & \cellcolor[HTML]{F8766D}0.07 & \cellcolor[HTML]{F8786E}0.08 \\ \hline

\end{tabular}%
}
\vspace{-5pt}
\end{table*}

\begin{table*}[!htbp]
\centering
\caption{Results across different country-selection settings (1st Split; 14-day lookback window).}
\vspace{-5pt}
\label{tab:integration_countries_1st_split}
\scriptsize
\renewcommand{\arraystretch}{1.15}

\resizebox{\textwidth}{!}{%
\begin{tabular}{llcccccccccccc}
\toprule
\multirow{3}{*}{} & \multirow{3}{*}{\makecell{Selection}}
& \multicolumn{12}{c}{1st Split} \\
\cmidrule(lr){3-14}
& &
\multicolumn{4}{c}{Overall Period (182 days)} &
\multicolumn{4}{c}{Non-wave Period (75 days)} &
\multicolumn{4}{c}{Wave Period (107 days)} \\
\cmidrule(lr){3-6}\cmidrule(lr){7-10}\cmidrule(lr){11-14}
& &
\makecell{MAE\\Avg} & \makecell{MAPE\\Avg (\%)} & \makecell{MAE\\Aggreg.} & \makecell{MAPE\\Aggreg. (\%)} &
\makecell{MAE\\Avg} & \makecell{MAPE\\Avg (\%)} & \makecell{MAE\\Aggreg.} & \makecell{MAPE\\Aggreg. (\%)} &
\makecell{MAE\\Avg} & \makecell{MAPE\\Avg (\%)} & \makecell{MAE\\Aggreg.} & \makecell{MAPE\\Aggreg. (\%)} \\
\midrule

\multirow{4}{*}{\makecell{Baseline\\Methods}} &
Na\"ive last      & 445 & 32.0 & 2472 & 24.3 & 173 & 34.6 & 1007 & 26.3 & 636 & 30.2 & 3499 & 22.8 \\
& Last week's avg & 492 & 35.4 & 3122 & 31.6 & 256 & 42.1 & 1695 & 36.7 & 657 & 30.6 & 4122 & 28.0 \\
& Weekly na\"ive  & 503 & 38.1 & 3122 & 31.6 & 256 & 44.7 & 1695 & 36.7 & 676 & 33.4 & 4122 & 28.0 \\
& ARIMA  & 435 & 31.2 & 2408 & 23.8 & 187 & 34.6 & 1166 & 26.9 & 609 & 28.8 & 3279 & 21.7 \\
\midrule

\multirow{4}{*}{XGBoost}
& National & 461 (9) & 27.4 (0.5) & 2671 (47) & 21.4 (0.3) & 137 (4) & 25.0 (0.6) & 756 (26) & 18.4 (0.5) & 688 (14) & 29.1 (0.7) & 4014 (66) & 23.5 (0.4) \\
& Correlated    & \textbf{352 (7)} & 24.4 (0.5) & \textbf{1925 (30)} & 18.9 (0.3) & 153 (4) & 26.3 (0.6) & 922 (20) & 20.7 (0.3) & 492 (11) & 23.0 (0.6) & 2629 (42) & 17.6 (0.3) \\
& Uncorrelated  & 423 (4) & 25.1 (0.3) & 2487 (14) & 19.4 (0.2) & 137 (3) & 25.2 (0.5) & 762 (15) & 18.1 (0.3) & 623 (7) & 25.1 (0.3) & 3696 (21) & 20.2 (0.1) \\
& All Countries & 361 (7) & 24.5 (0.3) & 2029 (21) & 18.7 (0.1) & 146 (3) & 25.9 (0.4) & 852 (9) & 19.4 (0.1) & 512 (10) & 23.5 (0.5) & 2853 (32) &  18.3 (0.2)\\
\midrule

\multirow{4}{*}{Transformer}
& National & 487 (109) & 32.4 (7.8) & 2854 (829) & 24.3 (7.4) & 160 (35) & 28.8 (4.2) & 889 (271) & 19.7 (4.1) & 716 (168) & 34.8 (10.9) & 4232 (1282) & 27.6 (10.2) \\
& Correlated    & 408 (156) & 26.5 (9.0) & 2382 (1211) & 21.1 (9.2) & 146 (35) & 25.5 (3.8) & 861 (256) & 19.6 (4.0) & 592 (242) & 27.3 (12.9) & 3449 (1897) & 22.1 (13.2) \\
& Uncorrelated  & 372 (42) & \textbf{23.5 (2.6)} & 2198 (321) & \textbf{18.4 (2.3)} & 125 (14) & 23.5 (1.8) & 697 (103) & 17.1 (1.6) & 544 (67) & 23.5 (3.4) & 3250 (526) & 19.3 (3.1) \\
& All Countries & 370 (70) & 24.1 (3.8) & 2113 (509) & 18.6 (3.6) & 133 (12) & 23.8 (1.6) & 770 (90) & 17.7 (1.5) & 535 (115) & 24.3 (6.0) & 3055 (842) & 19.2 (5.8) \\
\bottomrule
\end{tabular}%
}
\vspace{-5pt}
\end{table*}

\begin{table}[!htbp]
\centering
\caption{Baseline Methods vs. XGBoost vs. Transformer (2nd Split; 14-day lookback window).}
\vspace{-5pt}
\label{tab:models_comparison_split2}
\small
\renewcommand{\arraystretch}{1.15}

\resizebox{\columnwidth}{!}{%
\begin{tabular}{llcccc}
\toprule
\multirow{2}{*}{} & \multirow{2}{*}{Selection}
& \multicolumn{4}{c}{2nd Split} \\
\cmidrule(lr){3-6}
& &
\makecell{MAE\\Avg} & \makecell{MAPE\\Avg (\%)} & \makecell{MAE\\Aggreg.} & \makecell{MAPE\\Aggreg. (\%)} \\
\midrule
\multirow{4}{*}{\makecell{Baseline\\Methods}} &
Na\"ive last      & 69 & 32.0 & 324 & 20.1 \\
& Last week's avg & 47 & 22.1 & 294 & 18.3 \\
& Weekly na\"ive  & 63 & 29.0 & 294 & 18.3 \\
& ARIMA  & 61 & 27.6 & 254 & 15.8 \\
\midrule
\multirow{2}{*}{XGBoost} &
National & 53 (1) & 25.3 (0.4) & 247 (2) & 15.2 (0.2) \\
& All Countries   & \textbf{40 (0)} & \textbf{19.2 (0.2)} & \textbf{198 (1)} & \textbf{11.9 (0.1)} \\
\midrule
\multirow{2}{*}{Transformer} &
National & 62 (4) & 28.7 (2.7) & 304 (32) & 18.4 (2.1) \\
& All Countries   & 43 (2) & 19.6 (0.7) & 214 (23) & 13.5 (1.6) \\
\bottomrule
\end{tabular}%
}
\vspace{-5pt}
\end{table}

\begin{table}[!htbp]
\centering
\caption{Baseline Methods vs. XGBoost vs. Transformer (3rd Split; 14-day lookback window).}
\vspace{-5pt}
\label{tab:models_comparison_split3}
\small
\renewcommand{\arraystretch}{1.15}

\resizebox{\columnwidth}{!}{%
\begin{tabular}{llcccc}
\toprule
\multirow{2}{*}{} & \multirow{2}{*}{Selection}
& \multicolumn{4}{c}{3rd Split} \\
\cmidrule(lr){3-6}
& &
\makecell{MAE\\Avg} & \makecell{MAPE\\Avg (\%)} & \makecell{MAE\\Aggreg.} & \makecell{MAPE\\Aggreg. (\%)} \\
\midrule
\multirow{4}{*}{\makecell{Baseline\\Methods}} &
Na\"ive last      & 210 & 39.3 & 1078 & 25.1 \\
& Last week's avg & 155 & 24.5 & 910 & 18.0 \\
& Weekly na\"ive  & 193 & 35.4 & 910 & 18.0 \\
& ARIMA  & 171 & 30.9 & 805 & 17.7 \\
\midrule
\multirow{2}{*}{XGBoost} &
National       & 153 (3) & 25.1 (0.5) & 745 (14) & 15.7 (0.3)\\
& All Countries   & \textbf{116 (2)} & \textbf{20.3 (0.3)} & \textbf{594 (4)} & \textbf{13.8 (0.1)}\\
\midrule
\multirow{2}{*}{Transformer} &
National       & 176 (19) & 30.5 (1.9) & 873 (173) & 16.3 (1.9) \\
& All Countries   & 122 (10) & 21.4 (1.2) & 615 (83) & \textbf{13.8 (0.8)} \\
\bottomrule
\end{tabular}%
}
\vspace{-5pt}
\end{table}

Tables~\ref{tab:integration_countries_1st_split},~\ref{tab:models_comparison_split2}, and~\ref{tab:models_comparison_split3} show that incorporating multiple countries drastically improves results compared to using only one country. Table~\ref{tab:integration_countries_1st_split} further suggests that restricting the training set to either the \textit{correlated} or \textit{uncorrelated} groups does not consistently outperform the \textit{all countries} category. For \mbox{XGBoost}, the \textit{correlated} and \textit{all countries} settings achieve similar performance, while for the Transformer the \textit{uncorrelated} and \textit{all countries} settings are comparable. Two factors may explain this. First, the \textit{all countries} setting provides substantially more training samples, which is beneficial for data-hungry models. Second, classifying countries as \textit{uncorrelated} based on a single global correlation statistic does not imply an absence of useful signal, since similarities may occur elsewhere in time (e.g., shared wave shapes, growth/decay regimes, or delayed/advanced surges) and in different periods than those dominating the correlation calculation. As a result, excluding countries based solely on correlation can discard informative patterns that improve generalisation.

\subsection{Baseline Methods, XGBoost, and Transformer}
Finally, we compare XGBoost and the Transformer against the baseline methods under two country-selection settings, \textit{national (Cyprus only)} and \textit{all countries}, across the three train-test splits, using a 14-day lookback window. Overall, training on \textit{all countries} consistently improves the performance of both ML models across all three splits and yields lower errors than the baseline methods (Tables~\ref{tab:integration_countries_1st_split},~\ref{tab:models_comparison_split2}, and~\ref{tab:models_comparison_split3}). This pattern suggests that cross-country augmentation provides additional transferable signal, helping the models generalise beyond Cyprus-specific dynamics and improving robustness. Importantly, these gains are observed across both average and aggregate error metrics, indicating not only improved per-day accuracy but also stable performance when errors are accumulated over a full week.

\section{Conclusion}\label{sec:conclusion}
\vspace{-3.4pt}
COVID-19 highlighted the persistent risk posed by large-scale infectious-disease events and the need for stronger preparedness. Reliable short-term forecasts can support timely, evidence-based decisions by public health authorities and healthcare services, for example by anticipating surges, supporting staffing and capacity planning, and informing the timing and intensity of non-pharmaceutical interventions. Focusing on Cyprus, our results suggest that, across ML models, cross-country augmentation of the training data with additional European countries can improve predictive performance compared to training on Cyprus alone and to baseline methods, with a 14-day lookback window providing a strong and consistent operating point across settings.

Within the broader spectrum of infectious diseases, future work could extend the framework to forecast healthcare-burden indicators such as hospitalisations and intensive care unit (ICU) admissions. Further directions include evaluating the approach across multiple target countries, as well as incorporating contextual and exogenous information (e.g., policies, variant dynamics) to better capture cross-country heterogeneity and improve robustness under evolving conditions.

\vspace{-3.4pt}

\section{Ethics Statement}\label{sec:ethics}
\vspace{-3.4pt}
This study uses only aggregated, country-level, publicly available data and does not involve identifiable personal data. Therefore, formal ethics approval was not required.

\vspace{-3.4pt}

\def\IEEEbibitemsep{0pt plus .5pt}
\def\IEEEbibitemindent{0pt}
\bibliographystyle{IEEEtran}

\bibliography{refs_main}

\end{document}